%% file: main.tex
\documentclass[%
 aip,
 pop,
 amsmath,amssymb,
 reprint,%
]{revtex4-1}

\usepackage{graphicx}
\usepackage{dcolumn}
\usepackage{bm}

\usepackage[utf8]{inputenc}
\usepackage[T1]{fontenc}
\usepackage{mathptmx}
\usepackage{etoolbox}

\input{wenyinsty}

\makeatletter
\def\@email#1#2{%
 \endgroup
 \patchcmd{\titleblock@produce}
  {\frontmatter@RRAPformat}
  {\frontmatter@RRAPformat{\produce@RRAP{*#1\href{mailto:#2}{#2}}}\frontmatter@RRAPformat}
  {}{}
}%
\makeatother
\begin{document}

\preprint{AIP/123-QED}

\title[Functional perturbation theory under axisymmetry]{Functional perturbation theory under axisymmetry:\\
Simplified formulae and their uses for tokamaks }
\author{Wenyin Wei}
     \affiliation{
    Institute of Plasma Physics, Hefei Institutes of Physical Science, Chinese Academy of Sciences, Hefei 230031, People's Republic of China
    }
    \affiliation{
    University of Science and Technology of China, Hefei 230026, People's Republic of China
    }
    \affiliation{
    Forschungszentrum J\"{u}lich GmbH, Institute of Fusion Energy and Nuclear Waste Management – Plasma Physics, 52425 J\"{u}lich, Germany
    }

\author{Liang Liao}
    \affiliation{
    Institute of Plasma Physics, Hefei Institutes of Physical Science, Chinese Academy of Sciences, Hefei 230031, People's Republic of China
    }
    \affiliation{
    University of Science and Technology of China, Hefei 230026, People's Republic of China
    }
    
\author{Alexander Knieps}
    \affiliation{
    Forschungszentrum J\"{u}lich GmbH, Institute of Fusion Energy and Nuclear Waste Management – Plasma Physics, 52425 J\"{u}lich, Germany
    }
\author{Jiankun Hua}%
    \affiliation{
    Forschungszentrum J\"{u}lich GmbH, Institute of Fusion Energy and Nuclear Waste Management – Plasma Physics, 52425 J\"{u}lich, Germany
    }
    \affiliation{
    International Joint Research Laboratory of Magnetic Confinement Fusion and Plasma Physics, State Key Laboratory of Advanced Electromagnetic Engineering and Technology, School of Electrical and Electronic Engineering, Huazhong University of Science and Technology, Wuhan 430074, People’s Republic of China
    }
   
\author{Yunfeng Liang*$^{,}$}
    \email{y.liang@fz-juelich.de}
    \affiliation{
    Institute of Plasma Physics, Hefei Institutes of Physical Science, Chinese Academy of Sciences, Hefei 230031, People's Republic of China
    }
    \affiliation{
    Forschungszentrum J\"{u}lich GmbH, Institute of Fusion Energy and Nuclear Waste Management – Plasma Physics, 52425 J\"{u}lich, Germany
    }
 
\author{Shaocheng Liu*$^{,}$}
    \email{scliu@dhu.edu.cn}
    \affiliation{
    College of Physics, Donghua University, Shanghai 201620, China
    }

\date{\today}

\begin{abstract}
In strictly axisymmetric configurations of tokamaks, field-line tracing reduces from a three-dimensional ODE~(ordinary differential equation) system to a two-dimensional one, where Poincar\'e-Bendixson theorem applies and guarantees the nonexistence of chaos. The formulae of functional perturbation theory~(FPT) mostly simplify to compact closed-form expressions to allow the computation to finish instantly, which could improve and accelerate the existing plasma control systems by detangling the plasma dynamics from the magnetic topology change. 
FPT can conveniently calculate how the key geometric objects of magnetic topology:
\begin{itemize}
    \item the divertor X-point(s) and the magnetic axis~(which represent the X/O-cycles respectively), 
    \item the last closed flux surface ~(LCFS for short, which is the (un)stable manifolds of the divertor X-point for tokamaks when axisymmetry is guaranteed) 
    \item flux surfaces~(\textit{i.e.} invariant tori, where \textit{invariant} means the field lines on such a torus would never leave the torus)
\end{itemize}
change under perturbation. For example, when the divertor X-point shifts outwards, the LCFS there must expand accordingly, but not necessarily for other places of the LCFS, which could also contract, depending on the perturbation. FPT can not only facilitate adaptive control of plasma,
but also enable utilizing as much as possible space in the vacuum vessel by weakening the plasma-wall interaction~(PWI) via tuning the eigenvalues of $\calDPm$ of the divertor X-point(s), such that the field line connection lengths in the scrape-off layer~(SOL) are long enough to achieve detachment. Increasing flux expansion $f_x$ is another option for detachment and can also be facilitated by FPT. 

Apart from the edge, FPT can also benefit the understanding of the plasma core. Since the magnetic axis O-point would also shift under perturbation and the shift is known by FPT, the O-point can be controlled without full knowledge of the plasma response, which shall not significantly change the tendency. On the other hand, if the tokamak is equipped with comprehensive diagnostics to accurately measure the magnetic axis shift, one can then infer the plasma response magnetic field by comparing the field externally imposed and the field needed to cause the measured change, of which the difference shall come from the plasma response.

\end{abstract}
\maketitle


\section{Introduction}
\label{sec:intro}
Functional perturbation theory (FPT) is established as a powerful theoretical framework for analyzing how changes in the magnetic field $\mathbf{B}$ can shift or deform key structures such as orbits, X/O-cycles\cite{wei2024orbitshifts}, stable/unstable manifolds\cite{wei2024stable_manifolds_shifts}, and flux surfaces~(invariant tori)\cite{wei2024invariant_tori}.
While those references addressed general \emph{three-dimensional} (3D) toroidal devices (e.g.\ stellarators or perturbed tokamaks) and even apply to arbitrary finite-dimensional systems, the \emph{strict axisymmetry} of tokamaks makes many formulae of FPT significantly simplify to closed-forms without the need to integrate.

Even though numerous plasma control systems\cite{degrave2022, felici2014, galperti2024, kerboua_benlarbi2024, murari2024, seo2024, treutterer_asdex_2014, vu2019, tracey2024, portone1997, deranian2005, humphreys_high_2005, humphreys_integrated_2005, humphreys_development_2007, xiao_east_2008, xing_strike_2015} have been successfully developed on various magnetic confinement fusion~(MCF) machines, the magnetic topology itself has not been singled out and handled alone, which means that little attention is paid to the fact that the plasma dynamics is entangled with the magnetic topology change. Such entanglement naturally brings ignorance of the structure of the magnetic field itself. 

For example, researchers on plasma detachment have been accustomed to discussing $R$-$Z$ sections and have realized the importance of the field-line connection length $L_c$ because the long distance allows the power entering the SOL to be dissipated. Yet, very few notice that $L_c$ is tunable by adjusting the parallelism of nearby field-line trajectories to the divertor X-cycle, which is easy to ignore because this dimension is orthogonal to the $R$-$Z$ section.

This paper sets forth the \emph{axisymmetric} forms of FPT\cite{wei2024orbitshifts,wei2024stable_manifolds_shifts,wei2024invariant_tori}, replacing $\phi$-integrals directly by their analytical results. Specifically, this paper simplifies the formulae of FPT on 
\qquad\qquad\begin{enumerate}[label=(\roman*), leftmargin=4.5em]
    \item \textbf{the shifts of orbit and periodic orbit}, 
    \item \textbf{the shifts of stable/unstable manifold}, and
    \item \textbf{the deformation of invariant torus}
\end{enumerate} under perturbation, such that the implementation of FPT is much facilitated in tokamaks devoid of externally imposed or internally generated 3D fields. 

\section{Axisymmetric FPT: the toroidal progression simplifies}
\label{sec:axisymmetricFPT}

Field-line tracing along a vector field in standard 3D cylindrical coordinates obeys the following equation
\begin{align}
  \dot{  \vect{x} }_{\text{pol}} &=
  \frac{ R \vect{B}_{\text{pol}} }{ B_\phi } (\vect{x}_{\text{pol}}, \underbrace{\phi}_{ \mathclap{\substack{\text{removed} \\ \text{under axisymmetry} } } }),
  \label{eq:field_line_tracing}
\end{align}
where "toroidal" and "poloidal" are defined in the most natural way with the standard cylindrical basis $(\hat{\vect{e}}_R, \hat{\vect{e}}_Z, \hat{\vect{e}}_\phi)$:
\[
 \vect{B} = \begin{bmatrix}
\hat{\vect{e}}_R & \hat{\vect{e}}_Z & \hat{\vect{e}}_\phi
\end{bmatrix} \begin{bmatrix}
B_R \\ B_Z \\ B_\phi
\end{bmatrix} = \begin{bmatrix}
    1 & \hat{\vect{e}}_\phi
\end{bmatrix}
\begin{bmatrix}
\Bpol \\ B_\phi
\end{bmatrix},
\]
and a field-line tracing trajectory is simply denoted by $\Xpol(\vectx_{0,\text{pol}},\phis,\phie)$, with $\vectx_{0,\text{pol}}$ the initiating point $(R,Z)$-coordinates, $\phis$ the starting angle and $\phie$ the ending angle.

\begin{figure*}[p]
  \centering
  \rotatebox{90}{%
    \begin{minipage}{1.1\textheight}
      \centering
      \includegraphics[width=0.95\paperheight]{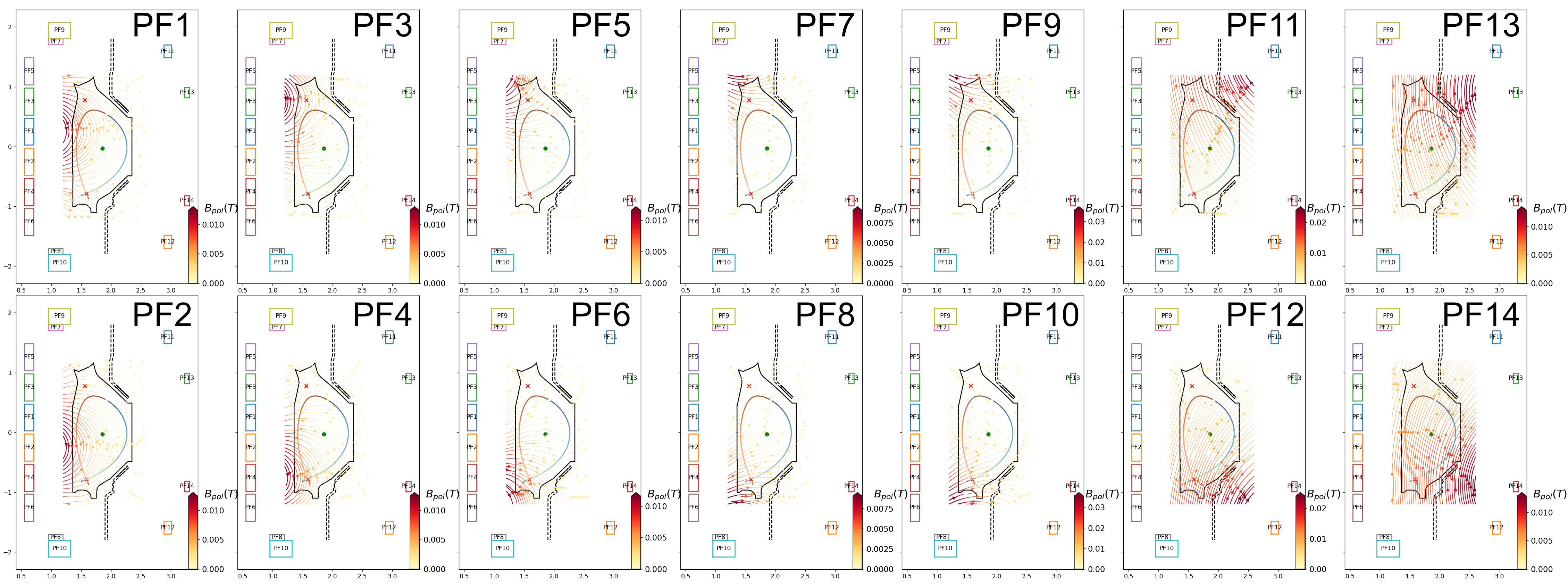}\par\vspace{1ex}
      \caption{\label{fig:EAST103950_PF} EAST PF coil vacuum fields per 1kA. The \#103950 LCFS at $t=3.0s$ and the first wall are drawn for reference. The magnetic field calculation is based on formulae given by [\onlinecite{labinac_magnetic_2006}].}
    \end{minipage}
  }
\end{figure*}

Convevntionally, for a rational flux surface with $q=m/n$, $m$ is the toroidal turn number of any cycle on this surface, while $n$ is that for poloidal. The definition of a periodic cycle can be generalized to those on irrational flux surfaces by enforcing $n=1$ and allowing $m$ to be an irrational number (\textit{i.e.} let $m$ take the value of $q$). Be aware that there does not exist a real cycle corresponding to $\calP^m$ when $m$ is irrational, but one can still regard $\calP^m$ as the returning map. Even if $m$ is irrational, the $\calDPm$ evolution formula\cite{wei2023} still works,
\begin{align}
    \frac{\rmd}{\rmd \phi} \calDPm = \left[\frac{\partial (R\Bpol / B_\phi) }{\partial (R,Z) }, ~~ \calDPm \right] = [\matr{A}, ~~ \calDPm ]
\end{align}
where $\matr{A}$ is $\partial (R\Bpol/B_\phi) / \partial (R,Z)$ abbreviated. 

When the vector field itself is considered as a function argument of the trajectory, the field-line tracing equation is complicated into 
\begin{align}
    \frac{\partial }{\partial \phie} \Xpol[\calB](\vectx_{0,\text{pol}}, \phis, \phie)
    = \frac{ R\Bpol }{B_\phi} [\calB](\overbrace{\Xpol}^{ \mathclap{ \Xpol[\calB](\vectx_{0,\text{pol}}, \phis, \phie) } }, \phie)
\end{align}
and, with the total functional derivative exerted on both sides, converts to the progression formula\cite{wei2024orbitshifts} of first-variation $\delta\Xpol$ under perturbation,
\begin{align}
    \frac{\partial }{\partial \phie}  \delta \vect{X}_\text{pol}
    = \delta (\frac{ R\Bpol }{B_\phi} )
+  \left( \delta \vect{X}_\text{pol} \cdot \frac{\partial}{\partial (R,Z)} \right) \frac{ R\Bpol }{B_\phi} .
\label{eq:progression_deltaXpol}
\end{align}
 
For tokamaks, the poloidal field~(PF) coils play an essential role in controlling the magnetic topology. Please refer to Fig.~\ref{fig:EAST103950_PF} for the vacuum field generated by the EAST PF coil system. The magnetic field calculation is based on formulae given by [\onlinecite{labinac_magnetic_2006}]. On the other hand, toroidal field~(TF) coils are usually operated with a fixed value of currents for tokamaks.

\subsection{X/O-cycles}
For X/O-cycles, the $\phi$-integration is trivial because their $(R,Z)$-coordinates are fixed due to axisymmetry, \textit{e.g.} $A:=\partial(R\mathbf{B}_{\mathrm{pol}} / B_\phi)/\partial(R,Z)$ is now constant, independent of $\phi$. Hence, the progression formula of $\calD\Xpol$ reduces to a constant-coefficient ODE:
\begin{equation}
   \frac{\partial}{\partial \phie}\calD\Xpol(\vectx_{0,\text{pol} }, \phis, \phie) 
   =\matr{A}\cdot\calD\Xpol(\vectx_{0,\text{pol} }, \phis, \phie),
\end{equation}
of which the initial condition is $\calD\Xpol(\vectx_{0,\text{pol} }, \phis,\phie)|_{\phie=\phis}=\mathrm{I}.$ The solution at $\phie=\phis\pm 2\pi$ can be immediately acquired in a form of matrix exponential, 

\begin{align}
    \calD\calP^{m=\pm 1} = \calD\Xpol|_{\phie=\phis+2\pi}
    &
    =e^{ 2\pi \matr{A}}=e^{2\pi \matr{A}[\calB](\xcyc[\calB])},
    \label{eq:XOcycle_DPm}
\end{align}
where $\pm$ takes the sign of  $B_\phi$, because if $B_\phi<0$ the field-line tracing is in fact backwards in $\phi$. 

Notably, $\calDPm$ of a cycle with fixed $(R,Z)$-coordinates must be constant no matter which $R$-$Z$ section is chosen to be the Poincar\'e $\calP$ section provided $\vect{B}$ is axisymmetric. This can also be checked by the $\calDPm$ evolution formula because the changing rate of 
$\calDPm$ \textit{w.r.t.} $\phi$ simply vanishes under axisymmetry, as shown below, 
\begin{align*}
  \frac{\mathrm{d} }{\mathrm{d} \phi }\calD\calP^{m=\pm 1} 
  &= \left[
        \frac{\partial (R\vect{B}_{\text{pol}}/B_\phi )  }{\partial (R,Z)}, 
    \calDPm
  \right] = \left[ \matr{A}, \calDPm \right] \\
  &= \left[
    \matr{A}, e^{2\pi \matr{A}}
  \right] 
\text{\quad ($\matr{A}$ and $e^{2\pi\matr{A}}$ obviously commute)} \\
&=  \matr{0} .
\end{align*}

The deduction of the cycle shift relies on $\calDPm$. Since the computation of $\calDPm$ can be significantly reduced by replacing the $\phie$-integral directly with the integral in a form of matrix exponential, the cycle shift can be described by the following concise formula (proof in Appendix \ref{ap:high_order_X_O}) 
\begin{align}
    \delta \xcyc &= -  \matr{A}^{-1}    \cdot \delta ( \frac{R \Bpol }{B_\phi } ),
    \\
    \textit{i.e. }
\delta \xcyc [\calB; \Delta \calB]
    &= -  \matr{A}^{-1} [\calB](\xcyc)   \cdot \delta ( \frac{R \Bpol }{B_\phi } )[\calB;\Delta\calB](\xcyc)
    \nonumber, 
\end{align}
where the latter equation with no label is the same as the former one but with all arguments including $\calB$ explicitly stated.

With $\delta\xcyc$ prepared, one can further investigate how $\calDPm$ of this cycle changes under perturbation. There is a strong motivation to estimate whether a perturbation would push $\calDPm$ eigenvalues away from or closer to unity, because this is relevant to how strongly the currents of poloidal field coils should adjust to keep $L_c$ in the divertor region long enough for detachment. The estimation is made easy and accurate with the aid of functional derivative, one simply needs to impose the total directional functional derivative $(\Delta\calB\cdot\frac{\rmd}{\rmd\calB})$ on both sides of Eq.~\eqref{eq:XOcycle_DPm},
\begin{align}
    & (\Delta\calB\cdot\frac{\rmd}{\rmd\calB})\calD\calP^{m=\pm 1} 
    \nonumber \\
    &= (\Delta\calB\cdot\frac{\rmd}{\rmd\calB})e^{2\pi \matr{A}}
    \nonumber \\
    &= \int_0^1 e^{\alpha \cdot 2\pi\matr{A}} \bigcdot (\Delta\calB\cdot\frac{\rmd}{\rmd\calB}) (2\pi\matr{A}) \bigcdot e^{(1-\alpha) 2\pi \matr{A} } \rmd \alpha
\end{align}
Since $\matr{A}$ here is evaluated as $\matr{A}[\calB](\xcyc[\calB])$, the total functional derivative consists of two components: 
\begin{align}
    &(\Delta\calB\cdot\frac{\rmd}{\rmd\calB}) (2\pi\matr{A}) \nonumber \\
    =& 
    \underbrace{
    (\Delta\calB\cdot\frac{\delta}{\delta\calB}) (2\pi\matr{A})
    }_\text{directly from the local field change}
    + \underbrace{
    \left(\delta\xcyc \cdot \frac{\partial}{\partial (R,Z)} \right) (2\pi\matr{A})
    }_\text{indirectly from the cycle shift via chain rule} ,
\end{align} 
based on which one can further discuss how the eigenvalues of $\calDPm$ of the divertor X-cycle(s) change under perturbation.

\begin{figure}[htbp]
\includegraphics[width=1.15\linewidth]{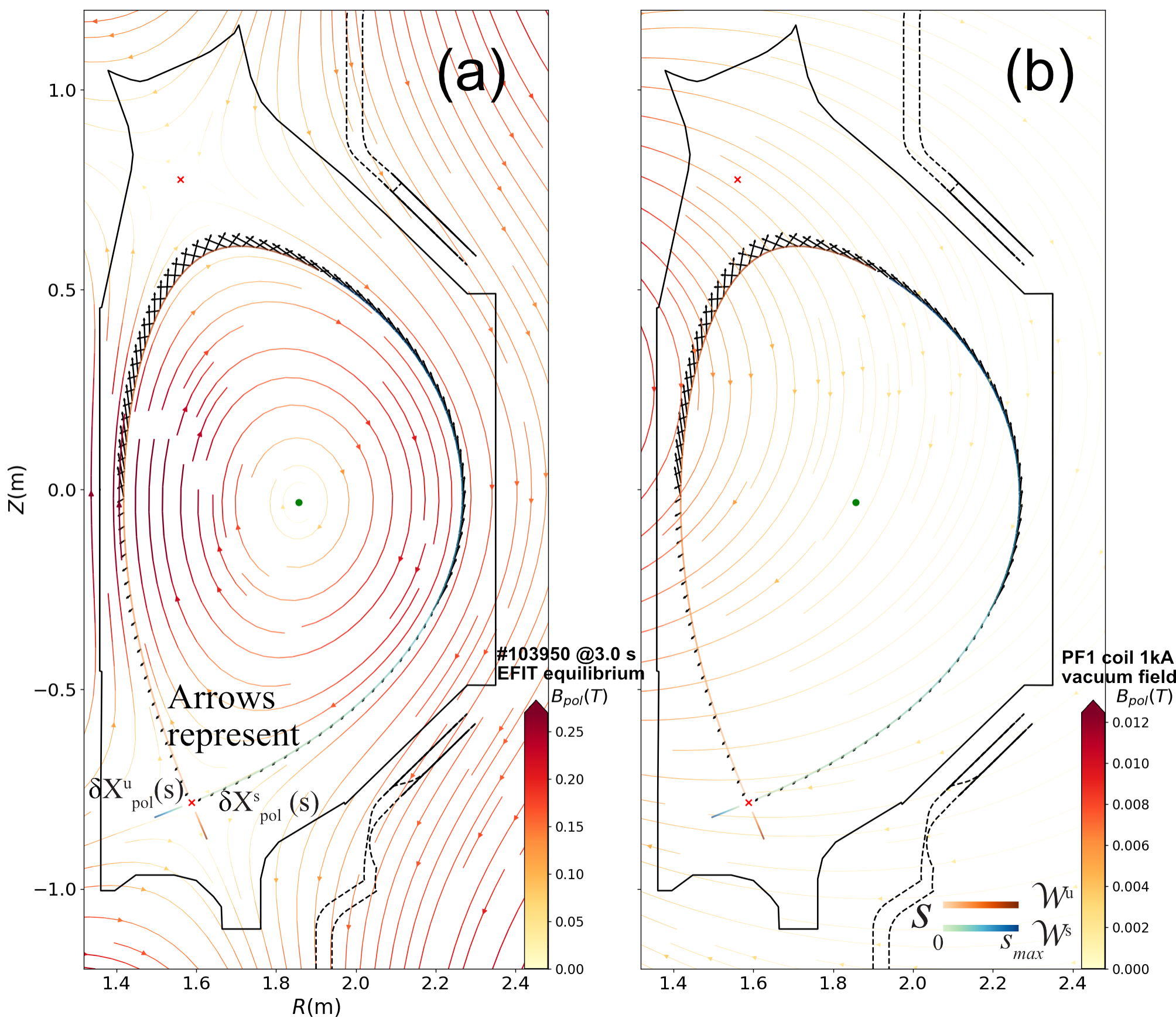}
\caption{\label{fig:EAST103950_LCFS_shift} Shifts $\delta\Xuspol(s)$ of the stable and unstable manifolds grown from the divertor X-point under the perturbation of 1kA PF1 coil vacuum field.  The streamline in (a) represents the EFIT poloidal field while that in (b) represents the vacuum field of PF1 coil with 1kA current. The shift $\delta\Xpol^\text{u}(s)$ progressed on the unstable branch shall have the same normal component as that of $\delta\Xpol^\text{s}(s)$ progressed on the stable branch, which is verified by the arrows in (a). The shift in the normal direction is an intrinsic property in terms of differential geometry, unlike the shift in the tangent direction (which can be affected by the parameterization). The first variations $\delta\Xuspol(s)$ here are progressed by Eq.~\eqref{eq:deltaXuspol_progression}. }
\end{figure}
\section{Stable and Unstable Manifolds}
\label{sec:stable_manifold}

Under axisymmetry, the invariant manifold growth formula in cylindrical coordinates\cite{wei2023} 
\begin{align}
    \partial_s \Xuspol (s,\phi)
    &= \frac{
        R\Bpol / B_\phi - \overbrace{ \partial_\phi \Xuspol }^{\mathrlap{ = \vect{0} \text{ under axisymmetry} } }    
    }{ \underbrace{\pm  | R\Bpol / B_\phi - \partial_\phi \Xuspol |}_{ =\rmd s / \rmd \phi } } 
\label{eq:invariant_manifold_growth}  \\
    &= \underbrace{\pm}_{ \mathrlap{ := \operatorname{sgn}(\rmd s/ \rmd \phi) } }\hatb_\text{pol} / \operatorname{sgn} (B_\phi) 
\\
    &= \underbrace{\pm}_{ \mathrlap{ \substack{ := +1 \text{for unstable manifolds},\\ -1 \text{ for stable manifolds} } } }\hatb_\text{pol}
\label{eq:invariant_manifold_growth_axisymmetric} 
\end{align}
simply reduces to the unit vector $\hatb_\text{pol}$~(or its reverse) along the polodial field $\Bpol=[B_R, B_Z]^\text{T}$. The sign $\pm$ in the formula above is the sign of $\rmd s / \rmd \phi$. For field-line tracing as $\phi$ increases, if the 1D (un)stable manifold arc length $s$ increases, $\rmd s / \rmd \phi$ takes a positive sign, then the sign $\pm$ takes $+$, otherwise $-$. Care must be taken to notice $s$ is the 1D (un)stable manifold arc length (measured at each $R$-$Z$ section), not the arc length along a field line.

Thereafter, one can recover the function argument $\calB$ and impose functional derivatives on both sides to acquire the first variation as below,
\begin{align}
    \partial_s \Xuspol[\calB] (s) & =\pm  \hatb_\text{pol}[\calB](\Xuspol)  
\\
    &= 
    \pm
    \begin{bmatrix} 
        B_R \\ B_Z
    \end{bmatrix} / \sqrt{B_R^2 + B_Z^2}
    \nonumber
\\
    \delta \partial_s \Xuspol[\calB;\Delta\calB] (s) &  
    \label{eq:deltaXuspol_progression}
\\
    = \pm & \Bigg\{ 
    \delta\hatb_\text{pol} + \Big(\delta\Xuspol\cdot \frac{\partial}{\partial (R,Z)} \Big)\hatb_\text{pol} 
    \Bigg\}
    \nonumber
\\
    =\pm & \Bigg\{ 
    \overbrace{
    \begin{bmatrix} 
        B_Z^2 \delta B_R - B_R B_Z \delta B_Z\\ 
        -B_R B_Z \delta B_R + B_R^2 \delta B_Z
    \end{bmatrix} / (B_R^2 + B_Z^2)^{3/2}
    }^{ \delta\hatb_\text{pol} := }
    \label{eq:Wus_growth_s_parameter}
\\
    & + \Big(\delta\Xuspol\cdot \frac{\partial}{\partial (R,Z)} \Big)\hatb_\text{pol} [\calB]
    \left(\Xuspol[\calB] (s) \right)
    \Bigg\}
    \nonumber
\end{align}
Notice that the parameterization based on the arc length $s$ causes difficulty in deducing high-order formulae, because the term $\hatb_\text{pol}=\vect{B}/|\vect{B}|$ is a normalized vector, where the denominator $|\vect{B}|$ makes it inconvenient to do differentiation.  

Obviously, $\hatb_\text{pol}$ is also the direction toward which a flux surface grows poloidally. Using another parameter $\zeta$ to replace the arc length $s$ allows the high-order formulae to manifest in a pretty concise form,
\begin{align}
    \partial_\zeta \Xuspol[\calB] (\zeta) &= \pm \Bpol[\calB](\Xuspol)  
\\    
    \delta\partial_\zeta \Xuspol[\calB] (\zeta) &= \pm \left\{ \delta\Bpol + \delta\Xuspol\cdot\frac{ \partial \Bpol }{ \partial (R,Z) } \right\}
\end{align}
where $\zeta:=\int\frac{1}{|\Bpol|}\rmd s$ is a modified arc length, as opposed to the standard arc length $s$. In other words, the modified arc length differential $\rmd\zeta= \frac{1}{|\Bpol|}\rmd s$ is the key to remove the normalisation of $\hatb_\text{pol}$. 

The $\zeta$ parameterization has almost the same $n$-th order formulae expressions as the $s$ one, as shown below. As mentioned before, the $s$ parameterization has the normalization issue of $\hatb_\text{pol}$ that impedes $\delta\hatb_\text{pol}$ to be conveniently expressed. Notably, the $\zeta$ parameterization also has its singularity issue, which will be discussed later.
\begin{widetext}
\begin{align}
    \frac{1}{n!} \delta^n \partial_s \Xuspol(s) 
= \sum_{\substack{ 
    \left( n_{\text{pol}, i} \right),
    \left( p_{\text{pol}, i} \right),
    n_{\hatb_{\text{pol}}} \\
    \text{such that} \\
    \sum_{i=1}^{d_{\text{pol} }}  n_{\text{pol}, i} p_{\text{pol}, i}
    + n_{\hatb_{\text{pol}}} = n
 } } 
    \binom{ p^{+}_{\text{pol} } }{ p_{\text{pol}, 1}, \dots, p_{\text{pol}, d_\text{pol}  } }
    ( \frac{  \delta^{n_{\text{pol}, 1} } \Xuspol  }{ n_{\text{pol}, 1} ! } )^{p_{\text{pol}, 1}}
    ( \frac{  \delta^{n_{\text{pol}, d_\text{pol} } } \Xuspol  }{ n_{\text{pol}, d_\text{pol} } ! } )^{p_{\text{pol}, d_\text{pol} }}
    \cdddot_{(p^+_\text{pol} )} 
    \frac{
    \partial_{(R,Z)}^{p^+_\text{pol} }
    \delta^{n_{\hatb_{\text{pol}}} } \hatb_{\text{pol}}
    }{ p^+_{\text{pol} }!  n_{\hatb_{\text{pol}}}! } ,
\\
    \frac{1}{n!} \delta^n \partial_\zeta \Xuspol(\zeta) 
= \sum_{\substack{ 
    \left( n_{\text{pol}, i} \right),
    \left( p_{\text{pol}, i} \right),
    n_{\Bpol} \\
    \text{such that} \\
    \sum_{i=1}^{d_{\text{pol} }}  n_{\text{pol}, i} p_{\text{pol}, i}
    + n_{\Bpol} = n
 } } 
    \binom{ p^{+}_{\text{pol} } }{ p_{\text{pol}, 1}, \dots, p_{\text{pol}, d_\text{pol}  } }
    ( \frac{  \delta^{n_{\text{pol}, 1} } \Xuspol  }{ n_{\text{pol}, 1} ! } )^{p_{\text{pol}, 1}}
    ( \frac{  \delta^{n_{\text{pol}, d_\text{pol} } } \Xuspol  }{ n_{\text{pol}, d_\text{pol} } ! } )^{p_{\text{pol}, d_\text{pol} }}
    \cdddot_{(p^+_\text{pol} )} 
    \frac{
    \partial_{(R,Z)}^{p^+_\text{pol} }
    \delta^{n_{\Bpol} } \Bpol
    }{ p^+_{\text{pol} }!  n_{\Bpol}! } ,
\end{align}
\end{widetext}
where 
\begin{align*} 
&  n_{\hatb_{\text{pol}}} \geq 0, \quad n_{\Bpol} \geq 0  \\
&  p^+ = p_1 + p_2 + \cdots + p_d, \quad d \text{ is the total number of powers,}\\
&  p_i \geq 1, \\ 
&  n_1 > n_2 > \cdots > n_d \geq 1,
\end{align*}
with subscripts $_\text{pol}$  of $(n_{\text{pol},i}, p_{\text{pol},i}, p^+_{\text{pol} }, d_{\text{pol} })$ omitted for brevity. 

Notice that the $\zeta$ parameterization based on might not be suitable to grow a \textit{stable/manifold} due to that the $\frac{1}{|\Bpol|}$ magnitude is proportional to $1/s$ near an X-point, hence 
\begin{align*}
\zeta_e - \zeta_s = \int_s^e \rmd \zeta= \int_\text{s}^\text{e} \frac{1}{|\Bpol|} \rmd s \sim \int_\text{s}^\text{e} s^{-1}\rmd s = \ln s|_\text{s}^\text{e} 
\\
\text{ (the $  _\text{s}^\text{e} $ sub/superscripts mean \textbf{s}tart/\textbf{e}nd.) }  
\end{align*}
is infinite when the starting point has arc length $s=0$, inducing inconvenience for parameterization.  Nevertheless, the $\zeta$ parameterization could be utilized to investigate the deformation of \textit{flux surfaces~(center manifolds)}.

\section{Flux Surface Deformation}
\label{sec:flux_surface_deform}

To investigate the deformation of a flux surface, one can set up the initial condition of $\delta\partial_\zeta \Xuspol = \delta \Bpol$ to be 
\begin{itemize}
    \item $\delta \Xuspol|_{\zeta=0}=\delta\xcyc$, where $\delta\xcyc$ is the cycle shift (hence it is implied that the torus is identified by its rotation transform),
    \item or $\delta \Xuspol|_{\zeta=0}=\vect{0}$ which implies the torus is anchored by a fixed point (during the dynamic process of perturbation, the torus passing the point $\Xuspol|_{\zeta=0}$ would always be considered as the object of concern).
\end{itemize}

Yet, for the nested closed flux surfaces in an MCF machine, a much more common parameterization is based on PEST flux coordinates $(\theta,r)$, with azimuthal angle $\phi$ omitted for axisymmetry. Therefore, we turn to simplify the flux surface deformation formulae from Ref.~\onlinecite{wei2024invariant_tori}. 

In tokamaks, usually one draws a horizontal line from the magnetic axis to the low-field side, on which the points are defined to have poloidal angles $\theta=0$. To facilitate setting such a condition, one can consider $(\theta, r)$ more fundamental than the state vector $\vectx$ representing $(R,Z)$-coordinates for tokamaks, \textit{i.e.}
\begin{center}
 let $\vectx$ be a function of $(\theta, r)$, that is $\vect\rchi(\theta, r)$. 
\end{center}
As a curved version of $\vect{X}$, the Greek letter $\vect\rchi$ is chosen here to represent the state vector but of a flux surface. In [\onlinecite{wei2024invariant_tori}], both $\vectx$-based and $(\theta,r)$-based invariant torus deformation formulae are provided. The $(\theta,r)$-based formulae shall be easier to put into practice for MCF due to the aforementioned reason, some of which are repeated below from [\onlinecite{wei2024invariant_tori}] and then simplified via axisymmetry. 

First of all, according to the definition of flux coordinates, the Poincar\'e map $\calP$ increments the PEST poloidal angle each time by a fixed value $\Delta\theta$, hence mapping repeated $k$ times means
\begin{align}
\calP^k \left(\vect\rchi(\theta, r)\right) := \vect\rchi(\theta + k \Delta\theta, r),    
\label{eq:Pk_defining}
\end{align}
of which the Jacobian is 
\begin{align}
\calD\calP^k ( \vect\rchi(\theta, r) )
&= \left.\begin{bmatrix}
  | \\ k\frac{\rmd \Delta\theta}{ \rmd r } \partial_\theta \vect\rchi   \\ | \end{bmatrix}\right|_{ \mathrlap{\smash{ \begin{subarray}{l} 
\theta + k  \Delta\theta  
\end{subarray} } } }   \cdot \begin{bmatrix}
    - ~ \nabla r ~ -
\end{bmatrix} \Big|_{ \begin{subarray}{l} \theta  \end{subarray} } 
\nonumber \\
& + 
\left. \begin{bmatrix}
 | &  | \\
 \partial_r \vect\rchi & \partial_\theta \vect\rchi \\
 | & | 
\end{bmatrix} \right|_{ \mathrlap{\theta+k\Delta\theta} }
\cdot
\left. \begin{bmatrix}
    - ~ \nabla r ~ - \\
    - ~ \nabla \theta ~ -
\end{bmatrix} \right|_{ \mathrlap{\theta} },
\label{eq:DPk_by_theta_grid}
\intertext{where the vertical and horizontal lines could be helpful to tell how the vectors fill the matrix, whether as a column or a row. The subscripts $_{\theta}$ and $_{\theta+k\Delta\theta}$ indicate where to evaluate the functions. A special case of $\calD\calP^k$ is when $k$ takes the value of $m$ (the latter term on the RHS degrades to an identity matrix $\matr{I}$) such that $\calP^m$ is a returning map for the points on this torus corresponding to $m$, }
\calD\calP^m ( \vectx )
&= \begin{bmatrix}
  | \\ m\frac{\rmd \Delta\theta}{ \rmd r } \partial_\theta \vect\rchi   \\ | \end{bmatrix}  \begin{bmatrix}
    - ~ \nabla r ~ -
\end{bmatrix}
 +  \matr{I}_{N\times N}
\label{eq:DPm_by_theta_grid}
\end{align}
The subscripts $_{\theta}$ and $_{\theta+k\Delta\theta}$ no longer appear because now they indicate the same place to evaluate, so there shall be no confusion.

Thanks to the axisymmetry of tokamaks, it is convenient to calculate the $\Delta\theta$, safety factor $q$, or rotation transform $\iota$ of a flux surface. One simply needs to trace a field line by the most basic FLT Eq.~\eqref{eq:field_line_tracing} and record when it returns to a point having the same $(R,Z)$-coordinates as its initiating point~(not necessarily all the three $(R,Z,\phi)$ coordinates have to be same). The recorded toroidal angle span, $\phie-\phis$, corresponds to $2\pi$ poloidal angle change, hence $q = \rmd\phi/\rmd \theta = (\phie-\phis) / 2\pi = 1/\iota$ whereas $\Delta\theta= (2\pi)^2 / (\phie-\phis)$. 

Hereafter, the concept of functional is introduced. If all arguments including the map $\calP$ itself are stated explicitly, the defining equation~\eqref{eq:Pk_defining} for $\calP^k$ becomes
\begin{align}
    \calP^k [\calP]\left(
    \vect\rchi[\calP](\theta,r)
    \right)
    = \vect\rchi[\calP](\theta+k\Delta\theta[\calP](r), r),
    \label{eq:Pk_defining_fundamental_theta_r}
\end{align}
which after imposed $\Delta\calP\cdot\rmd / \rmd \calP$ converts to
\begin{align}
    \delta\calP^k ( \vect\rchi(\theta,r))
    + \calD\calP^k ( \vect\rchi(\theta,r))
    \cdot  \delta\vect\rchi ( \theta,r)
\nonumber \\
    =  \delta\vect\rchi (\theta + k\Delta\theta,r)
    + \left.\Big(\partial_\theta \vect\rchi\right)\Big|_{\mathrlap{\smash{
        (\theta + k\Delta\theta, r) 
    } } } 
    \cdot \overbrace{
        \delta(k\Delta\theta)
    }^{\mathclap{\text{vanishes if $\Delta\theta$ is merely dependent on $r$, not on $\calP$.}}},
    \label{eq:deformation_formula_second_form}
\end{align}
that is the $(\theta,r)$-based form of \textit{the first-order flux surface deformation formula under perturbation}. If the flux surface label $r$ is chosen to be $\Delta\theta$, $q$ or $\iota$, then $\Delta\theta$ is naturally only dependent on $r$, totally independent of $\calP$. Nonetheless, if the flux surface label $r$ is chosen to be the toroidal flux, the poloidal flux or the volume enclosed or similar effective radius, $\Delta\theta=\Delta\theta[\calP](r)$ would have to be a function with both $\calP$ and $r$ as arguments. Only choices like $r:=\Delta\theta$, $q$ and $\iota$ can make $\Delta\theta = \Delta\theta[r]$ a pure function of $r$. 

Solving the flux surface deformation formula~\eqref{eq:deformation_formula_second_form} for the deformation $\delta\vect\rchi$ is much facilitated under axisymmetry. Firstly, one can not only calculate $\delta\calP^k$ for each integer $k$ by the progression formula~\eqref{eq:progression_deltaXpol} of first variation $\delta\Xpol$ under perturbation, but also for each real number $k$ by the fact that toroidal tracing for $2\pi$ angle is equilvalent to a unity increment in $k$ (this only works under axisymmetry). Secondly, calculate $\calD\calP^k$ by Eq.~\eqref{eq:DPm_by_theta_grid} according to the $(\theta,r)$ grid that has been prepared before. Thirdly, notice that $\delta\vect\rchi$ is a $2\pi$-periodic function in $\theta$, hence one can fit its Fourier coefficients by least squares.

For a high-order version of flux surface deformation formula under perturbation, please refer to Appendix A of [\onlinecite{wei2024invariant_tori}], which can be simplified if $\delta(k\Delta\theta)$ vanishes when the torus label is chosen to be $r:=\Delta\theta$, $q$ or $\iota$.

The condition that the points having the same $Z$-coordinate at the low-field side as that of the magnetic axis are defined to have $\theta=0$ can be easily expressed in the following $(\theta,r)$-based form,
\begin{align}
    &\hat{\vect{e}}_Z \cdot \delta^n\vect\rchi(\theta =0, r) 
    = \underbrace{ \hat{\vect{e}}_Z \cdot \delta^n \vect\rchi(\theta=0, r=0) }_{\mathclap{\text{ the magnetic axis shift $Z$-component} }}, ~~  n\geq 0
\end{align}
where $\delta^n\vect\rchi$ is simply the $n$-th order variation under perturbation.

\section{Conclusion and discussion}
We have outlined how the general functional perturbation theory (FPT) from Refs.~\cite{wei2024orbitshifts,wei2024stable_manifolds_shifts,wei2024invariant_tori} drastically simplifies under strict axisymmetry, making the formulae simpler and suitable to be incorporated by plasma control systems. Computation can be near-instant for typical tokamak operations, thereby enabling real-time fine-tuning of magnetic topology for tokamaks.

When the $\calDPm$ eigenvalues of the divertor X-point are tuned close enough to unity, the higher-order terms $\calD^k\calP^m$ with $k>=2$ can play an important role in determining the local behaviour of nearby field-line trajectories. To calculate these terms, one can utilize the progression equations in Appendix~\ref{ap:high_order_sensitivity}.


\appendix

\section{The shift of an X/O-point under axisymmetry}
\label{ap:high_order_X_O}

First of all, calculate the shift of the trajectory corresponding to the X/O-point at $\phie=\phis+2\pi$ (without loss of generality, assume $B_\phi>0$). The shift is induced by the perturbation (also axisymmetric) and accumulates in the integration process from $\phis$ to $\phie$, as described by the following progression formula of $\delta\Xpol$:
\begin{align*}
\frac{\partial }{\partial \phie}  \delta \vect{X}_\text{pol}(\vectx_{0,\text{pol}}, \phis, \phie)
&= \left( \delta \vect{X}_\text{pol} \cdot \frac{\partial}{\partial (R,Z)} \right) \frac{ R\Bpol }{B_\phi} 
+  \delta (\frac{ R\Bpol }{B_\phi} )
\\
&= \matr{A}\cdot\delta\Xpol + \delta(\frac{ R\Bpol }{B_\phi} ) 
\tag*{\eqref{eq:progression_deltaXpol} revisited}
\end{align*}
Notice both $\matr{A}$ and $\delta(R\Bpol / B_\phi)$ are constant for X/O-points under axisymmetry. The constant-coefficient nonhomogeneous ODE system has the following solution at $\phie=\phis+2\pi$:
\begin{align}
\delta\calP &:= 
\left. \delta \Xpol \right|_{\phie = \phis + 2\pi}
\nonumber\\
&= e^{ \matr{A} (\phie - \phis) }\cdot \vect{0} + \int_{\phis}^{\phie} e ^{\matr{A}(\phie -\phi) } \cdot \delta ( \frac{R \Bpol }{B_\phi } ) ~ \rmd\phi
\nonumber \\
&= e^{\matr{A}\phie} \cdot \int_{\phis}^{\phie} e ^{- \matr{A} \phi }  \rmd\phi \cdot \delta ( \frac{R \Bpol }{B_\phi } )
\nonumber \\
&= e^{\matr{A}\phie} \cdot 
(-\matr{A}^{-1}) \cdot \left. e ^{- \matr{A} \phi } \right|_{\phis}^{\phie}   \cdot \delta ( \frac{R \Bpol }{B_\phi } )
\nonumber \\
&= e^{\matr{A}\phie} \cdot 
(-\matr{A}^{-1}) \cdot (e^{- \matr{A} \phie } - e^{ - \matr{A} \phis})   \cdot \delta ( \frac{R \Bpol }{B_\phi } )
\nonumber \\
&= 
(-\matr{A}^{-1}) \cdot (\matr{I} - e^{  \matr{A} (\phie - \phis)})   \cdot \delta ( \frac{R \Bpol }{B_\phi } )
\nonumber \\
&= 
\matr{A}^{-1} \cdot (e^{  2\pi\matr{A} } - \matr{I} )   \cdot \delta ( \frac{R \Bpol }{B_\phi } )
\end{align}

To investigate the cycle shift rather than the ending point shift of a trajectory initiating at a fixed point, the starting point needs to move to keep matched with the ending point. Additionally, this starting point movement would propagate to the ending point via $\calDPm$. As concluded in [\onlinecite{wei2024orbitshifts}], the cycle shift obeys the equation $\delta\xcyc = - (\calDPm-\matr{I})^{-1}\cdot \delta\calP$. With $\delta\calP$ prepared above, the cycle shift has a concise formula to compute under axisymmetry, as shown below,
\begin{align}
    \delta \xcyc &= - (\calDPm- \matr{I})^{-1} \cdot \delta \calP 
    \nonumber \\
    &= - (e^{2\pi\matr{A} }- \matr{I})^{-1} \cdot \delta \calP 
    \nonumber \\
    &= - (e^{2\pi\matr{A} }- \matr{I})^{-1} \cdot \matr{A}^{-1} \cdot (e^{  2\pi\matr{A} } - \matr{I} )   \cdot \delta ( \frac{R \Bpol }{B_\phi } ),
    \nonumber \\
    &= -  \matr{A}^{-1}    \cdot \delta ( \frac{R \Bpol }{B_\phi } ),
\end{align}
which is simple enough for real-time computation.

\section{Progression equations of high-order sensitivity tensors}
\label{ap:high_order_sensitivity}
The sensitivity of a field-line trajectory $\Xpol$ to initial condition can be tracked by various orders of $\calD^k \calP^m$, which can be acquired by integrating the following progression equations of different $k$-orders,
\allowdisplaybreaks
\begin{align}
\text{(0th)} \frac{\partial}{\partial \phie} \Xpol( \vectx_{0, \text{pol}}, \phis, &\phie) = 
    \frac{R\Bpol}{B_\phi} (\Xpol( \vectx_{0, \text{pol}}, \phis, \phie ), \phie)
\\
\text{(1st)} \frac{\partial}{\partial \phie}\calD\Xpol &= 
     \calD\Xpol \cdot\frac{ \partial (R\Bpol / B_\phi)  }{\partial (R,Z) }
\\
\text{(2nd)} \frac{\partial}{\partial \phie}\calD^2\Xpol &= 
    \calD^2\Xpol \cdot \frac{ \partial (R\Bpol / B_\phi)  }{\partial (R,Z) }
\\
    & + \calD\Xpol \calD\Xpol \cddot \frac{ \partial^2 (R\Bpol / B_\phi)  }{\partial (R,Z)^2 }
\nonumber \\
\text{(3rd)} \frac{\partial}{\partial \phie}\calD^3\Xpol &= 
    \calD^3\Xpol \cdot \frac{ \partial (R\Bpol / B_\phi)  }{\partial (R,Z) }
\\
\indices{_{d_3}_{d_2}_{d_1}_x } &\qquad \indices{ _{d_3}_{d_2}_{d_1}_i } \qquad \indices{^i_x } 
\nonumber\\
    & + \calD\Xpol \calD^2\Xpol \cddot \frac{ \partial^2 (R\Bpol / B_\phi)  }{\partial (R,Z)^2 }
\nonumber\\
&\qquad \indices{ _{d_3}_j }\quad \indices{ _{d_2}_{d_1}_i }  \qquad \indices{^j^i_x } 
\nonumber\\
    & + \calD^2\Xpol \calD\Xpol \cddot \frac{ \partial^2 (R\Bpol / B_\phi)  }{\partial (R,Z)^2 }
\nonumber\\
&\qquad \indices{ _{d_3}_{d_2}_j }\quad \indices{ _{d_1}_i }  \qquad \indices{^j^i_x } 
\nonumber\\
    & + \calD\Xpol \calD^2\Xpol \cddot \frac{ \partial^2 (R\Bpol / B_\phi)  }{\partial (R,Z)^2 }
\nonumber\\
&\qquad \indices{ _{d_2}_j }\quad \indices{ _{d_3}_{d_1}_i }  \qquad \indices{^j^i_x } 
\nonumber\\
    & + \calD\Xpol \calD\Xpol \calD\Xpol \cdddot \frac{ \partial^3 (R\Bpol / B_\phi)  }{\partial (R,Z)^3 }
\nonumber  \\  
&\qquad \indices{ _{d_3}_k }\quad \indices{ _{d_2}_j } \quad \indices{ _{d_1}_i }  \qquad\quad \indices{^k^j^i_x } ,
\nonumber
\end{align}
where Einstein convention is used and indices are marked below tensors to clarify which two dimensions are imposed dot product.

Poincar\'e mapping is simply the trajectory $\vect{X}(\vectx_{0,\text{pol}}, \phis, \phie)$ when $\phie = \phis + 2m\pi$ for a cycle of $m$ toroidal turns. The mapping can be expanded as Taylor series as below,
\begin{align}
    \calP^m (\vectx_{0,\text{pol}})
    & = \overbrace{ \calP^m (\xcyc) }^{\mathrlap{ =\xcyc \text{ by definition} }}
\\
    & + \Delta\vectx_{0,\text{pol}} \cdot \calD\calP^m (\xcyc) 
\nonumber \\
    & + \Delta\vectx_{0,\text{pol}}\Delta\vectx_{0,\text{pol}} \cddot \calD^2\calP^m (\xcyc) 
\nonumber \\
    & + \Delta\vectx_{0,\text{pol}}\Delta\vectx_{0,\text{pol}}\Delta\vectx_{0,\text{pol}} \cdddot \calD^3\calP^m(\xcyc) 
\nonumber \\
    & + \mathcal{O}(|\Delta \vectx_{0,\text{pol}} |^4 )
\nonumber
\end{align}
where $\Delta\vectx_{0,\text{pol}}$ is $\vectx_{0,\text{pol}} - \xcyc$. When $\calDPm(\xcyc)$ degrades to the identity matrix, the role of the higher-order terms begins to assert itself. A true snowflake divertor configuration is only possible with a degraded $\calDPm(\xcyc)$, and it would have more than four X-legs. The figurative noun X-leg can be made a strict definition: \textit{invariant branch}. Let $\gamma$ be a hyperbolic cycle and $\mathcal{W}^\text{u}(\gamma)$ the stable manifold approaching the cycle $\gamma$. An invariant branch of $\mathcal{W}^\text{u}(\gamma)$ is defined here to be a connected component of $\mathcal{W}^\text{u}(\gamma)\setminus\gamma$. By removing $\gamma$ from the manifold $\mathcal{W}^\text{u}(\gamma)$, there appears two branches~(connected components) if $\calDPm$ does not degrade.

\bibliography{main}

\end{document}

%% file: wenyinsty.tex
\usepackage[cal=pxtx]{mathalpha}
\usepackage{enumitem} 
\usepackage{tensor}

\usepackage{rotating} 

\newcommand{\vect}[1]{\mathbf{\boldsymbol{#1}}} 
\newcommand{\matr}[1]{\mathbf{#1}} 

\newcommand{\calD}{\mathcal{D}}
\newcommand{\calB}{\mathcal{B}}

\newcommand{\calP}{\mathcal{P}}

\newcommand{\calDPm}{\mathcal{DP}^{m}}

\newcommand{\Xuspol}{ \vect{X}^{\text{u/s}}_{\text{pol}} }
\newcommand{\phis}{ \phi_{\text{s}} }
\newcommand{\phie}{ \phi_{\text{e}} }
\newcommand{\Xpol}{ \vect{X}_{\text{pol}} }
\newcommand{\Bpol}{ \vect{B}_{\text{pol}} }

\newcommand{\rmd}{\mathrm{d}}
\newcommand{\hatb}{\hat{\vect{b}} }

\newcommand{\xcyc}{\vect{x}_{\text{cyc}} }

\newcommand{\vectx}{ \vect{x} }

\DeclareRobustCommand{\rchi}{{\mathpalette\irchi\relax}}
\newcommand{\irchi}[2]{\raisebox{\depth}{$#1\chi$}} 

\usepackage{mathtools} 

\usepackage{scalerel}
\usepackage{stmaryrd}

\def\onedot{$\mathsurround0pt\ldotp$}
\def\cddot{
  \mathbin{\vcenter{\baselineskip.67ex
    \hbox{\onedot}\hbox{\onedot}}%
  }}%
\def\cdddot{
  \mathbin{\vcenter{\baselineskip.67ex
    \hbox{\onedot}\hbox{\onedot}\hbox{\onedot}%
  }}%
}

\usepackage{graphicx}
\makeatletter
\newcommand*\bigcdot{ {\mathpalette\bigcdot@{.5}} }
\newcommand*\bigcdot@[2]{\mathbin{\vcenter{\hbox{\scalebox{#2}{$\m@th#1\bullet$}}}}}
\makeatother

\usepackage{tikz}

\tikzset{down square arrow/.style={to path={-- ++(0,-.25) -| (\tikztotarget)}}}
\tikzset{up square arrow/.style={to path={-- ++(0,+.25) -| (\tikztotarget)}}}
\tikzset{
    up square arrow cdot/.style={
        to path={-- ++(0,0.25) -| (\tikztotarget) node[pos=0.25, fill=white] {$\cdot$}}
    },
    down square arrow cdot/.style={
        to path={-- ++(0,-0.45) -| (\tikztotarget) node[pos=0.25, fill=white] {$\cdot$}}
    }
}

\usepackage{cancel}